\documentstyle[prb,aps,multicol,epsf]{revtex}
\begin{document}
\draft
\title{Theory of spin wave excitations of metallic A-type antiferromagnetic manganites}
\author{G. Jackeli$^{a,b,}$\cite{byline},
N.B. Perkins$^{a}$,
and N.M. Plakida$^{a}$}
\address{
$^{a}$Joint Institute for Nuclear Research, Dubna, Moscow region,
141980, Russia \\
$^{b}$
Max-Planck-Institut f$\ddot{o}$r
Physik komplexer Systeme,
N$\ddot{o}$thnitzer Str. 38
01187 Dresden, Germany.}
\maketitle
\begin{abstract}\widetext
The spin dynamics of the metallic A-type antiferromagnetic manganites is
studied. An effective nearest-neighbour Heisenberg spin wave dispersion
is derived from the double exchange model taking into account the
superexchange interaction between the core spins. The result of inelastic
neutron scattering experiment on ${\rm Nd}_{0.45}{\rm Sr}_{0.55}{\rm Mn}
{\rm O}_{3}$ is qualitatively reproduced. Comparing theory with
experimental data  two main parameters of the model: nearest-neighbour
electron transfer amplitude and superexchange coupling between the core
spins are estimated.
\end{abstract}
\begin{multicols}{2}
\narrowtext
\newpage

Interest in the doped manganese oxides with perovskite structure
${\rm R}_{1-x}{\rm B}_x{\rm MnO}_3$ (where
${\rm R}$ is trivalent rare--earth and ${\rm B}$ is divalent alkaline ion,
respectively) has revived since the discovery of colossal magnetoresistance
(CMR) in those compounds.\cite{cmr}

Extensive experimental study of the doped manganites has
revealed  unusual microscopical structures of spin--, charge--, and
orbital--ordered states.\cite{rev}
At different doping concentration
a full range of magnetically ordered states  such as
antiferromagnetic (AFM) insulator, ferromagnetic (FM)
metal and  charge ordered (CO) insulator have been  observed.
The importance of the charge, spin as well as orbital\cite{nag} and lattice
\cite{mil} degrees of freedom and of interplay between them has now been well
established.

In the ideal undistorted perovskite structure ${\rm Mn}$
ions are surrounded by
six oxygen ions, and  form a simple cubic lattice with the trivalent rare-earth
ion sitting in the center of the cube.
The six oxygen ions, ${\rm O}^{2-}$, forming the octahedra give rise to the crystal
field potential that removes the degeneracy of $d$ orbital of manganese.
As a result $d$ level is splitted into $t_{2g}$  triplet
($d_{xy}$, $d_{xz}$, and
$d_{yz}$) and $e_{g}$ doublet ($d_{3z^2-r^2}$, and $d_{x^2-y^2}$).
The  orbitals $d_{3z^2-r^2}$, and $d_{x^2-y^2}$ are degenerate and since they
point toward the  ${\rm O}^{2-}$ ions they have higher energy then the $t_{2g}$
orbitals. The $t_{2g}$ electrons
are strongly coupled by intra-atomic Hund's coupling in
the high spin state $S=3/2$, and remain localized even in doped compounds.
The $e_{g}$ electrons are  subject to strong Hund's coupling with the core
spin, they are localized in the undoped compounds (one $e_{g}$ electron per
manganese ion) due to the strong intra-atomic Coulomb repulsion.

The reference undoped compound ${\rm La}{\rm Mn}{\rm O}_{3}$ is a
Mott insulator with a spin $S=2$, and orbital degrees of freedom.
Spins are ordered ferromagnetically in $(x,y)$ plane and
antiferromagnetically in $z$ direction (so-called A-type AFM state).\cite{wol,GOOD} The orbitals are also ordered:
in-plane directional orbitals $d_{3x^2-r^2}$ and $d_{3y^2-r^2}$ alternate
in $(x,y)$ plane and are stacked in $z$ direction.
This leads to the  distortion of local surrounding of the Jahn-Teller (JT)
active ${\rm Mn}$ ion, rezulting in   elongation of ${\rm Mn}{\rm O}_{6}$ octahedra along $x$
($y$) direction which has been observed experimentally.\cite{jt}

Upon doping of the parent compounds the $e_g$-electrons become mobile
using  oxygen's  $p$-orbitals as a bridge between ${\rm
Mn}$ ions and the strong on-site Hund's coupling drives core spins to align
parallel forming the metallic FM ground state.
The double exchange (DE) interaction\cite{de} that mediates the ferromagnetic coupling
between the core spins is the key ingredient in a appropriate interpretation
of the  phase diagram at the doping range $0.2<x<0.5$ where FM metallic
behavior is observed.

Close to the half-doping ($x\sim 0.5$) the phase diagram becomes more
subtle. The narrow band manganites (low $T_{\rm c}$ compounds) exhibit a
charge ordered state close to the commensurate filling. This CO state is
characterized by an alternating ${\rm Mn}^{3+}$ and ${\rm Mn}^{4}$ ions
arrangement in the $(x,y)$ plane with charge stacking in $z$-direction.
In addition, these systems show $d_{3x^2-r^2}$/$d_{3y^2-r^2}$ orbital
ordering. In CO state these systems exhibit an insulating behavior with a
very peculiar form of AFM spin ordering at low temperature. The observed
magnetic structure is a  CE-type and consists of quasi one-dimensional
ferromagnetic zig-zag chains coupled antiferromagnetically.\cite{wol} A
direct evidence of the CE charge/spin  ordered state in half-doped
manganites has been provided by the electron diffraction for ${\rm
La}_{0.5}{\rm Ca}_{0.5}{\rm MnO}_3$.\cite{cheong} Similar observations
have also been reported for ${\rm Pr}_{0.5}{\rm Ca}_{0.5}{\rm
MnO}_3$.\cite{kaj,zim} All these compounds show  paramagnetic insulating
states at high temperature with predominant FM fluctuations.\cite{kaj}
Upon lowering the temperature the  charge correlations together with
orbital correlation develops. 
The FM fluctuations are first  suppressed at
charge/orbital ordering temperature $T_{\rm o}$ and completely disappear at 
 the AFM ordering temperature $T_{\rm N}$.\cite{kaj} And finally, at $T<T_{\rm
N}$ a CE charge/spin/orbital ordered state is established.

It is clear that  all the degrees of freedom involved in the formation of
ground state  are strongly coupled at commensurate doping.
Then  the symmetry of
the ground state is dictated by the competition between the
hopping driven ferromagnetic exchange, antiferromagnetic superexchange (SE),
electrostatic and lattice elastic energies. This competition has been studied
in detail  to interpret the rather complicated CE spin/charge/orbital
ordered state observed in half-doped manganites.\cite{GOOD,CE,US}

In the moderately narrow band system
 ${\rm Nd}_{0.5}{\rm Sr}_{0.5}{\rm MnO}_3$\cite{kuw} and
${\rm Pr}_{0.5}{\rm Sr}_{0.5}{\rm MnO}_3$\cite{PrSrCO1} CO state has been
observed in very narrow band region of ${\rm Sr}$
concentration around $x\sim 0.5$.

However, in  a recent experimental study of ${\rm Pr}_{0.5}{\rm
Sr}_{0.5}{\rm MnO}_3$ and ${\rm Nd}_{0.45}{\rm Sr}_{0.55}{\rm MnO}_3$ it was found\cite{PrSrNCO} that
instead of the well known CE charge/spin/orbital  ordered state both systems
show A-type AFM order in the ground state with uniform
$d_{x^2-y^2}$ orbital order and no clear sign of the  charge
ordering. The A-type AFM order has been also confirmed by the large
anisotropy in resistivity  and spin wave measurements.\cite{AAFM}

This finding is consistent with recently obtained mean--field phase digram of
two--orbital double--exchange model.\cite{US} It was shown that
A-type AFM state and charge ordering do not coexist.
The A-type spin ordering is unstable against the formation
CE-type magnetic structure
when charge ordering  is introduced in the system.
In the CO state the carrier kinetic energy and hence double exchange energy are
suppressed, and the CE spin/charge  ordered
state is more favorable due to the additional gain in energy by modulating the
FM bonds in basal plane and generating a "dimerization"--like gap
at the Fermi surface due to a staggered factor in electron hopping
amplitude.

In the present paper we propose the   spin wave theory of such an A-type
metallic state.  We show that the spin wave excitation spectrum derived
in the leading order of $1/S$ expansion of the double exchange model in
A-type metallic state reproduces the spin wave dispersion relation
observed recently in the inelastic neutron scattering experiment on ${\rm
Nd}_{0.45}{\rm Sr}_{0.55}{\rm MnO}_3$.\cite{AAFM} Fitting theory to the
experimental data gives an estimate of effective transfer integral, as
well as the AFM superexchange coupling between the $t_{2g}$ core spins.

In the A-type AFM state  electrons are
confined in the basal plane due to the large Hund's coupling.
Since the transfer integral between
$d_{x^2-y^2}$ orbitals is  largest  in the $(x,y)$ plane, the low
energy eigen state of two orbital tight--binding Hamiltonian has mainly
$d_{x^2-y^2}$ character. Hence there exists an anisotropy in orbital
occupation, even in the absence of JT coupling. Including JT effect
 will lead to a farther increase of the anisotropy and  the splitting
between  $d_{x^2-y^2}$ and  $d_{3z^2-r^2}$ levels. Therefore, in our
model   we retain  only the relevant $d_{x^2-y^2}$ orbital, and also
assume that all the other degrees of freedom are integrated out to give
 effective model parameters. We start with the Hamiltonian:
\begin{eqnarray}
H&=&-t\sum_{\langle ij\rangle\parallel,\sigma}
\left[d_{i\sigma }^{\dagger}d_{j\sigma}+H.c.\right]-\mu\sum_{i}n_{i}\nonumber\\
&-&
J_{\rm H}\sum_{i}{\bf S}_i{\bf \sigma}_i
+J\sum_{\langle ij\rangle}{\bf S}_{i}{\bf S}_{j}.
\label{H1}
\end{eqnarray}
The first term  in Eq.(\ref{H1}) describes an electron hopping between
the the  nearest neighbor (NN) Mn-ions in $(x, y)$ plane, with $t$ being
the transfer amplitude between  $d_{x^{2}-y^{2}}$ orbitals. The second
term  describes the Hund's coupling between the spins of
localized $t_{2g}$-~electrons ${\bf S}_i$  and the itinerant $e_{g}$
electrons with spin ${\bf  \sigma}_{i}$. The  superexchange 
interaction  of localized spins between the NN sites is given by $J$,
$n_{i}$  is the particle number operator and $\mu$ is the chemical
potential.

Next step is to introduce two sublattices $A$ and $B$ with spin up and
down for the alternating in $z$-direction plains in A-type AFM state. 
Then we can expand the core spin operators by the Holstein-Primakoff
transformation as follows:
\begin{displaymath}
S_{i}^{+}=\left\{\begin{array}{ll}
\sqrt{2S}\alpha_{i}&\textrm{if $i\in A$}\\
\sqrt{2S}\beta_{i}^{\dagger}&\textrm{if $i\in B$}\\
\end{array}\right.~,\;\;
S_{i}^{-}=\left\{\begin{array}{ll}
\sqrt{2S}\alpha_{i}^{\dagger}&\textrm{if $i\in A$}\\
\sqrt{2S}\beta_{i}&\textrm{if $i\in B$}\\
\end{array}\right.~,
\end{displaymath}
\begin{eqnarray}
S_{i}^{z}=\left\{\begin{array}{ll}
S-\alpha_{i}^{\dagger}\alpha_{i}&\textrm{if $i\in A$}\\
-S+\beta_{i}^{\dagger}\beta{i}&\textrm{if $i\in B$}\\
\end{array}\right.~.
\label{HP}
\end{eqnarray}

By substituting expressions (\ref{HP}) into the Hamiltonian (\ref{H1})
and keeping the terms contributing to leading order in $1/S$ expansion
we obtain the following Hamiltonian in the momentum space
$
H=H_{\text{el}}+H_{\text{sw}}+H_{\text{int}},
$
where $H_{\text{el}}$ and $H_{\text{sw}}$ describe the charge and
spin degrees of freedom, respectively and interaction among them are
given by $H_{\text{int}}$:
\begin{eqnarray}
H_{\text{el}}&=&\sum _{{\bf k}\sigma} \epsilon_{{\bf k}\sigma}\left[
a_{{\bf k}\sigma}^{\dagger}a_{{\bf k}\sigma}+b_{{\bf
k}\bar{\sigma}}^{\dagger}b_{{\bf k}\bar{\sigma}}\right],
\label{H3}\\
H_{\text{sw}}&=&\sum _{\bf q}\left[\omega_{0,{\bf q}}\left(\alpha_{\bf
q}^{\dagger}\alpha_{\bf q}+\beta_{\bf q}^{\dagger}\beta_{\bf q}\right)+
\left(\Omega_{{\bf q}}\alpha_{\bf
q}^{\dagger}\beta_{-{\bf q}}^{\dagger}+\text{H.c.}
\right)\right],
\nonumber\\
H_{\text{int}}&=&-\frac{J_{\rm H}\sqrt{S}}{\sqrt{2N}}\sum_{{\bf k,\bf q}}
\left[a_{{\bf k}\uparrow}^{\dagger}a_{{\bf k}+{\bf q}\downarrow}\alpha_{\bf
q}^{\dagger}+b_{{\bf k}\downarrow}^{\dagger}b_{{\bf k}+{\bf q}\uparrow}
\beta_{\bf q}^{\dagger}+\text{H.c.}\right]
\nonumber
\end{eqnarray}
where $a_{{\bf k}\sigma}$ and $b_{{\bf k}\sigma}$ stands for  the electrons
on  $A$ and $B$ sublattice, respectively,
$\epsilon_{{\bf k},\uparrow(\downarrow)}=-4t\gamma_{\parallel{\bf k}}
\mp J_{\text H}S/2-\mu$ is the electron dispersion   in $(x,y)$ plane
including Zeeman splitting, $\gamma_{\parallel{\bf k}}=1/2(\cos k_{x}+\cos
k_{y})$,  $\omega_{0,{\bf q}}=J_{\text{H}}m+2JS-4JS(1-\gamma_{\parallel{\bf
q}})$, $\Omega_{\bf q}=JS(1+e^{-2iq_{z}})$, and $m=1/2(n_{a,\uparrow}-n_{a,\downarrow})=-1/2(n_{b,\uparrow}-
n_{b,\downarrow})$ is the magnetization of electron subsystem.
In the state with fully polarized FM planes considered here
 $m=n/2$ with $n$ being the electron density.

The spin wave part of the Hamiltonian $H_{\text{sw}}$ (\ref{H3}) is
diagonalized by the standard Bogoliubov transformation
\begin{eqnarray}
\alpha_{\bf q}=e^{-iq_{z}}[u_{{\bf q}} \hat \alpha_{\bf q} +v_{{\bf q}}
\hat \beta^{\dagger}_{\bf -q}],\; \beta_{-{\bf q}}^{\dagger}=-v_{{\bf q}}
\hat \alpha_{\bf q} +u_{{\bf q}} \hat \beta^{\dagger}_{\bf -q} \label{bog}
\end{eqnarray}
bringing it into the form
\begin{equation}
H_{\text{sw}} = \sum_{\bf q} \omega_{{\bf q}}
\left[ \hat{\alpha}^{\dagger}_{\bf q}\hat{\alpha}_{\bf q}  +
 \hat{\beta}^{\dagger}_{\bf q}\hat{\beta}_{\bf q} \right].
\label{HSWD}
\end{equation}
The eigen--frequency $\omega_{{\bf q}}$ and the coherence factors are
given by
\begin{eqnarray}
u_{{\bf q}}&=&\frac{1}{\sqrt{2}}\left[1+\frac{\omega_{0,{\bf q}}}
{\omega_{{\bf q}}}\right]^{1/2},\; v_{{\bf q}}=\frac{1}{\sqrt{2}}
\left[1- \frac{\omega_{0,{\bf q}}}
{\omega_{{\bf q}}}\right]^{1/2},\nonumber\\
\omega_{{\bf q}}&=&\sqrt{\omega_{0,{\bf q}}^{2}-|\Omega_{{\bf q}}|^{2}}.
\label{uv}
\end{eqnarray}

Next we  introduce the Fourier transformed  retarded matrix Green
function (GF) for magnons
${\mathsf D}_{{\bf q},\omega}=
\ll {\mathsf A}_{{\bf q}}|{\mathsf A}_{{\bf q}}^{\dagger}\gg_{\omega}$
where, ${\mathsf A}_{{\bf q}}$
is the two--component operator and
$A_{{\bf q}}^{\dagger} = ( \alpha_{{\bf q}}^{\dagger}, \;
\beta_{- {\bf q}} )$. The components of  $\cal{D}_{{\bf q},\omega}$ are
however related among each-other by symmetry relations and hereafter we
consider the diagonal (normal) $(\cal D)$ and non-diagonal (anomalous)
$(\hat  {\cal D})$ components, expressing them in terms of the self-energy
operators
  via the Dyson-Beliaev equation
\begin{eqnarray}
{\cal D}_{{\bf q},\omega}&=&\frac{\omega+\omega_{\bf
q}+\Sigma_{{\bf q},-\omega}}{{D}_{{\bf q},{\omega}}}~,\;\;\;
\hat{ \cal D}_{{\bf q},\omega}=\frac{-\hat\Sigma_{{\bf
q},\omega}}{{ D}_{{\bf q},{\omega}}}~,
\label{DB}
\end{eqnarray}
where the following notations has been introduced:
\begin{eqnarray}
{D}_{{\bf q},{\omega}}&=&[\omega-{\cal A}_{{\bf q},\omega}]^2-
[\omega_{\bf q}+S_{{\bf q},\omega}^{-}][\omega_{\bf q}+S_{{\bf
q},\omega}^{+}]~, \label{DBnot} \\
{\cal A}_{{\bf q},\omega}&=&\frac{\Sigma_{{\bf q},\omega}-\Sigma_{{\bf
q},-\omega}}{2},
\;
{\cal S}_{{\bf q},\omega}^{-}=\frac{\Sigma_{{\bf q},\omega}+
\Sigma_{{\bf q},-\omega}-2\hat\Sigma_{{\bf
q},\omega}}{2},\nonumber\\
 {\cal S}_{{\bf q},\omega}^{+}&=&\frac{\Sigma_{{\bf
q},\omega}+ \Sigma_{{\bf q},-\omega}+2\hat\Sigma_{{\bf q},\omega}}{2}~.
\label{DB1}
\end{eqnarray}
Here  $\Sigma_{{\bf q},\omega}$ and $\hat\Sigma_{{\bf q},\omega}$ are
normal and anomalous components of magnons self-energy due to the
coupling with electrons,
 ${\cal A}_{{\bf q},\omega}$ is an antisymmetric function in $\omega$,
while  ${\cal S}_{{\bf q},\omega}$ and  $\hat\Sigma_{{\bf q},\omega}$ are
symmetric functions of $\omega$.

\begin{figure}
      \epsfysize=35mm
      \centerline{\epsffile{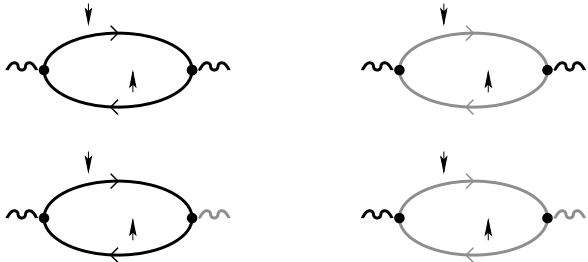}}
\caption{Graphical representation of the normal and anomalous components
of the spin wave self-energy. The black and gray straight  (wavy)  lines
stand for the  electron (spin wave)  propagators on $A$ and $B$ sublattices,
respectively.} \label{f1}
\end{figure}

In the leading order of $1/S$ expansion, the self-energy operators
($\Sigma$ and $\hat\Sigma$) are given by the particle-hole bubble
diagrams shown in Fig. 2. The black  and gray   lines  stand for the
electron propagators on $A$ and
 $B$ sublattices, respectively. The upper (lower) graphs corresponds to the
normal (anomalous) component of the magnons self-energy. The anomalous
part of the self-energy couples degenerate $\alpha$ and $\beta$ magnons
denoted by black and gray wavy lines, respectively. The dot represents
the magnon-fermion vertex, analytical form of which is easily obtained by
reexpressing the interaction term of the Hamiltonian (\ref{H3}) in terms
of transformed magnon operators. The above introduced dynamical quantities
(\ref{DBnot}) have the  following analytical expression:
\begin{eqnarray}
{\cal A}_{{\bf q},\omega}&=&\frac{J_{\text H}^2S}{2N}
\sum_{{\bf k}}
\frac{[n_{{\bf k}\uparrow}-n_{{\bf k + q}\downarrow}]\omega}
{\omega^2-[\epsilon_{{\bf k + q}\downarrow} -\epsilon_{{\bf k}\uparrow}
]^2}~,
\label{SE}\\
{\cal S}_{{\bf q},\omega}^{\pm}&=&\frac{J_{\text H}^2S
[u_{\bf q}\pm v_{\bf q}]^2}{2N}\sum_{{\bf k}}
\frac{[n_{{\bf k}\uparrow}-n_{{\bf k + q}\downarrow}][\epsilon_{{\bf k + q}\downarrow}-\epsilon_{{\bf k}\uparrow}
]}
{\omega^2-[\epsilon_{{\bf k + q}\downarrow}-\epsilon_{{\bf k}\uparrow}
]^2}~,
\nonumber
\end{eqnarray}
where $n_{{\bf k}\sigma}=[e^{\epsilon_{{\bf k}\sigma}}+1]^{-1}$ is the Fermi
distribution function.

The renormalized spin wave spectrum, arising from  the second order
self-energy corrections, are given by the poles of the magnons Green's
function (\ref{DB}),
or equivalently by the zeros of ${D}_{{\bf q},{\omega}}$ (\ref{DBnot}).
Assuming the large Hund's coupling $J_{\text H}\gg t$, and
expanding the self-energies up to leading order in $t/J_{H}$ the spin wave
spectrum can be written as:
\begin{eqnarray}
{\tilde \omega}_{{\bf q}}=2S\sqrt{[J+2J_{\text F}(1-\gamma_{\parallel{\bf
q}})]^{2}-J^2\gamma_{\perp{\bf
q}}^2}~,
\label{swsp}
\end{eqnarray}
where $\gamma_{\perp{\bf k}}=\cos k_{z}$ and 
the effective ferromagnetic exchange energy is given by
\begin{equation}
J_{\text F}=J_{\text{DE}}-J,
\;\; \text{with}\;\;  J_{\text{DE}}=t\langle
a_{i\uparrow}^{\dagger}a_{j\uparrow}\rangle/(2S^2)
\label{JDE}
\end{equation}
 being the double-exchange
energy given by kinetic energy
of electrons.

\begin{figure}
\epsfysize=80mm
\centerline{\epsffile{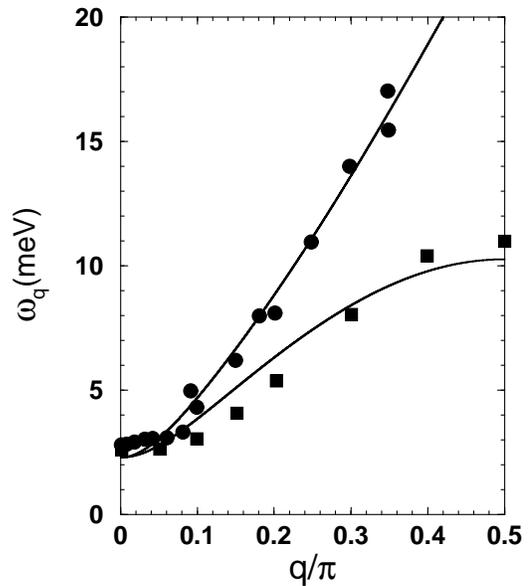}}
\caption{The spin wave dispersion of
${\rm Pr}_{0.5}{\rm Sr}_{0.5}{\rm Mn} {\rm O}_{3}$
from Ref.17 (the
full circles and squares stand for ${\bf q}\parallel (1,0,0)$ and ${\bf
q}\parallel (0,0,1)$, respectively). The solid curves represents fitting by
Eq.\ref{swsp} including the anisotropy gap.}
\label{f2}
\end{figure}

The obtained spin wave spectrum (\ref{swsp}) corresponds to that of
nearest-neighbor Heisenberg model with intraplane ferromagnetic exchange
integral $J_{\text F}$ and antiferromagnetic interplane coupling  $J$.
The correspondence between spin dynamics in the FM phase of DE model and
in effective NN Heisenberg model exist at quasi-classical level (i.e.
leading order in $1/S$ expansion) and has been studied in
Refs.\onlinecite{KUBO}.

The long wave-length behavior of the spin-wave dispersion is given by
\begin{eqnarray}
{\tilde \omega}_{{\bf q}}\propto\left\{\begin{array}{ll}
C_{\parallel}q&\textrm{for ${\bf q}\parallel (1,0,0)$}\\
C_{\perp}q&\textrm{for ${\bf q}\parallel (0,0,1)$}\\
\end{array}\right.
\label{limit}
\end{eqnarray}
where $C_{\parallel}=2S\sqrt{JJ_{\text{F}}}$ and $C_{\perp}=2SJ$ are intra--
and inter--plane spin wave stiffness, respectively.
Here we note, that long wave-length behavior of spin wave spectrum
within the FM planes reduces to sound-like dispersion unlike the $q^2$
behavior inherent for isotropic ferromagnets.

Fig. 2 shows a fit of inelastic neutron scattering experiment for ${\rm
Nd}_{0.45}{\rm Sr}_{0.55}{\rm Mn} {\rm O}_{3}$ from Ref.\onlinecite{AAFM}
using the  obtained spin wave spectrum (\ref{swsp})
including the phenomenological
anisotropy gap ${\Delta}$. 
The full circles and squares stand for the experimental
data in $(1,0,0)$ and $(0,0,1)$ direction respectively.
As has already been pointed out in Ref.\onlinecite{AAFM}, the best fit is
obtained for a  intralayer FM exchange $4J_{\text F}S\simeq 32\text{meV}$,
interlayer AFM exchange $2JS\simeq 10\text{meV}$, and
$\Delta\simeq 0.26\text{meV}$. At $x=0.55$ NN correlation function $\langle
a_{i\uparrow}^{\dagger}a_{j\uparrow}\rangle=0.2$ and from Eq.\ref{JDE} one obtains following
estimates for $d_{x^2-y^2}$ transfer integral $t\simeq 0.195\text{eV}$ and
AFM superexchange coupling $J\simeq 3.3\text{meV}$.

To summarize, the spin wave dispersion in A-type AFM state of the double
exchange model is derived tacking into account the superexchange interaction
between the core spins. The result of inelastic neutron scattering experiment
on ${\rm
Nd}_{0.45}{\rm Sr}_{0.55}{\rm Mn} {\rm O}_{3}$ is qualitatively reproduced.
Comparison of theory with experimental data gives the estimates of two main
parameters of the model: nearest-neighbor electron transfer amplitude
$t$ and superexchange coupling between the core spins
$J$. While the present estimate of AFM superexchange strength  $J\simeq
3.3\text{meV}$ coincides with that widely used in the literature,
\cite{j} it is factor 4
times larger then SE integral estimated form the Neel temperature of the
end compound ${\rm Ca MnO}_{3}$ leading to $J\simeq 0.8\text{meV}$.\cite{millis}
In ${\rm Sr}$ doped compounds the both $e_{g}$ and $t_{2g}$ ${\rm Mn}-{\rm O}$
transfer amplitudes ($t_{e}$ and $t_{g}$) are higher than that in ${\rm Ca}$ doped compounds due to
the different ionic radii of these element leading to different  ${\rm
Mn}-{\rm O}$ distance and ${\rm
Mn}-{\rm O}-{\rm Mn}$ bond angle. The AFM SE exchange is given by the
forth power of the transfer amplitude $t_{e}^{4}$ ($t_{g}^{4}$)
 and hence a small increase of the latter
might give a larger value of SE strength in Sr-doped compounds. This
increase of magnetic frustration due to the increase of AFM SE strength
going from ${\rm Sr}$ to ${\rm Ca}$ doped compounds has been also
observed experimentally~\cite{fr}.



The authors acknowledge the  kind hospitality at the
Max-Plank Institute f$\ddot{\rm u}$r Physik komplexer Systeme, where the
the present work has been carried out.
Financial support by the INTAS Program,
Grants No 97-0963 and No 97-11066, are also  acknowledged.
One of the authors (G.J.) is partly supported by
the programme SCOPES of the Swiss National Science Foundation
and  the Federal Department
of Foreign Affairs.
G.J. would like to thank Nic Shannon and Victor Yu. Yushankhai
for useful discussions.

\end{multicols}
\end{document}